# Parametric amplification of orbital angular momentum beams based on light-acoustic interaction

Wei Gao,[1,a,b] Chunyuan Mu,[1] Hongwei Li,[2] Yuqiang Yang[1], and Zhihan Zhu[2,a,b]

[1]*Institute of photonics and optical fiber technology, Harbin University of Science and Technology, Harbin 150080, China.*

[2]*The higher educational key laboratory for Measuring & Control Technology and Instrumentation of Heilongjiang Province, Harbin 150080, China.*

A high fidelity amplification of beams carrying orbital angular momentum (OAM) is very crucial for OAM multiplexing and other OAM-based applications. Here, we report the first study of stimulated Brillouin amplification (SBA) for OAM beams, the energy conversion efficiency of photon-phonon coupling and the phase structure of amplified signals are investigated in collinear and noncollinear frame systems, respectively. Our results demonstrate that the OAM signals can be efficiently amplified without obvious noise introduced, and the modes of output signal are independent of the pump modes or the geometrical frames. Meanwhile, an OAM state depending on the optical modes and the geometrical frames is loaded into phonons by coherent light-acoustic interaction, which reveals more fundamental significance and a great application potential in OAM-multiplexing.

Driven by a rapid growing demand for data volume in information explosion and big data era, mode division multiplexing, increasing both spectrum efficiency and information capacity density, has gradually developed.[1-3] Particularly, orbital angular momentum (OAM) modes have attracted considerable attention due to their unique feature of inherently infinite dimensionality,[4] and have a number of studies in high capacity information communication and processing based on OAM-multiplexing in recent years.[5-10] The essential units of OAM-multiplexing system can be summarized into two layers, i.e. transmission layer and manipulating layer, and an inevitable energy gap exists between the two layers due to their different physical ways.

Laser amplification is a straightforward method for compensating this gap, but an unacceptable noise level, which attenuates the OAM channels, embarrasses its implementation.[11] Second-order nonlinear optical processes, based on light-light interaction, provide flexible ways to manipulate the OAM states and also a feasible way to amplify OAM signals,[12-17] i.e. optical parametric amplification (OPA). However, OPA needs specific pump source, and is not well compatible with optical circuit. Brillouin photon-phonon coupling, a kind of third-order nonlinear optical process based on light-acoustic interaction, provides more freedoms of signal manipulating both in classical and quantum,[18-20] and can be well applied in fiber or on-chip integtated photonics.[7,20-21] In this letter, we demonstrate a mechanism for amplifying OAM beams with high efficiency and fidelity based on stimulated Brillouin ampilfication (SBA) process, involving collinear and noncollinear frames.

---

[a] Electronic mail: wei_g@163.com; zhuzhihandd@sina.com;
[b] Wei Gao and Zhihan Zhu contributed equally to this work.

The experimental setup is shown in Fig. 1. A frequency-doubled Q-switched Nd:YAG laser produces linear polarization Gaussian pulses with duration 12ns at 1Hz repetition rate. A half-wave plate and a polarized beam splitter (PBS$_1$) are used to control polarization and split the laser beam into two beams. The reflected beam is directed injected into the SBS cell contained CS$_2$. The returning Stokes beam passes through a SPP (RPC Photonics, VPP-1c) to form an OAM mode of $\ell = 1$. This mode is used as an input signal whose energy is varied by calibrated neutral-density filters (ND). The transmitted beam from PBS$_1$ is used to form a pump beam. The pump energy is continuously varied by means of an attenuator, which consisted of a rotatable half-wave plate and PBS$_2$. These two beams interact in the 30cm-long Brillouin amplifier (BA) contained CS$_2$, and amplified signal beam is detected by CCD or energy meter at the output position. The interference pattern of an OAM mode and a plane reference beam (red dotted line in Fig. 1(a)) can be recorded by the CCD to analyze the phase singularity of the output beam. The layout in the upper dashed box [see Fig. 1(a)] shows a collinear Brillouin amplification frame. Specifically, the vortex signal beam interacts with the counter-propagating pump beam in the BA cell. When the pump is injected at a small crossing angle ($\theta$) with respect to the propagating direction of the signal, the frame, as depicted in the lower dashed box of the Fig. 1(a), represents the noncollinear Brillouin amplification.

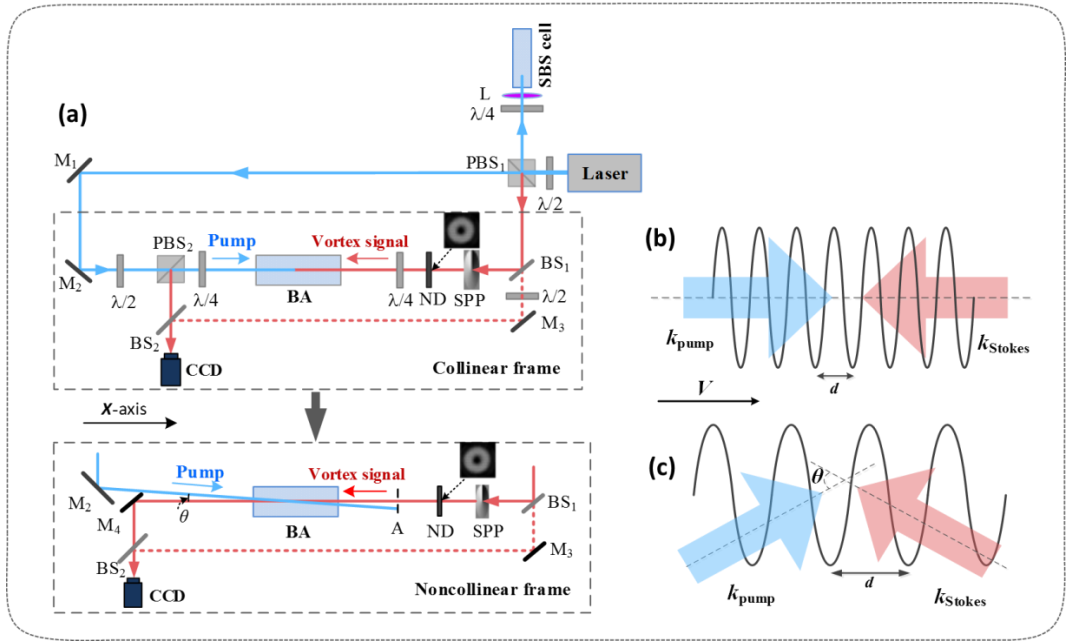

FIG. 1. (a) Experimental setup. Key components includes Brillouin amplifier (BA), spiral phase plate (SPP), polarized beam splitter (PBS); neutral-density filer (ND), aperture (A), mirrors (M), lens (L), beam splitter (BS). (b), (c) Schematic diagrams of the periodic optical potentials induced by pump and Stokes fields in collinear frame (b) and noncollinear frame (c), respectively.

We first discuss the OAM evolution law in the photon-phonon coupling process of SBA. For straightforward describing the relationship of energy flow and OAM in the interaction, a quantum description is used although all interacting fields are classical in this work. The interaction picture Hamiltonian of SBA and the OAM relationship between the boson fields are given by Eq. (1) and Eq. (2), respectively,

$$H = \hbar\kappa(ab^{\dagger}c^{\dagger} + a^{\dagger}bc) \tag{1}$$

$$L_x|\ell_a\rangle = L_x(|\ell_b\rangle + |\ell_c\rangle) \tag{2}$$

where $c$, $a$ ($b$) and $|\ell_c\rangle$, $|\ell_a\rangle$ ($|\ell_b\rangle$) are the boson annihilation operators for phonon and pump (signal) photon and their OAM states, respectively, and $\kappa$ is a coupling constant, $L_x$ is OAM operator for *X*-axis. The Eq. (1) and Eq. (2) imply that the SBA is a quasi-parametric down-conversion between photons and phonons in which signal field will be coherently enhanced without OAM modes changed or noise introduced. And we should note that although the SBA exhibits a theoretical zero noise characteristic, a mode noise introduced by self-SBS of pump is inevitable. Then one may naturally ask that: how does the system maintain the OAM conservation in the case of increasing the occupation number of OAM signal field. The answer to this question is that an OAM state depending on the OAM modes of pump and signal will be coherently transferred to phonon field, and this principle provides many exploitable resources for OAM-multiplexing in physical layer.

Then we discuss the effects of phase mismatching in noncollinear SBA process. In the collinear frame, as shown in Fig. 1(b), the counter-propagating pump and Stokes fields induce a 1D moving periodic optical potential at a velocity

$$v = d\Delta\Omega \tag{3}$$

where $v$ and $\Delta\Omega$ are the sound velocity and the Brillouin frequency shift of the media, respectively, and $d$ is "lattice constant" of the periodic potential depending on the wavelengths of pump and signal. In the noncollinear frame, as shown in Fig. 1(c), the $d$ will increase as the crossing angle $\theta$ increases according to a relationship of $d(\theta) = d/\cos(\theta/2)$. The increasing of $d$ means phase mismatching, however, according to Eq. (3), the phase mismatching caused by crossing angles can be compensated by means of decreasing the frequency shift between pump and signal, and this proposal have been experimentally verified by a previous work.[22] Notice that the effect of phase mismatching is negligible for $\theta$ less than 10°;

therefore, there is no need to take the compensation in situation of quasi-collinear frame adopted in this work. In addition, it is important to note that even though the $\theta$ changing a lot or pump carrying OAM, the amplified signal will still keep OAM constant for the reason that the system OAM remains conservative at the cost of changing OAM of excited phonon. More specific details are out of the scope of this letter and are covered in a separate article.

Now we experimentally investigate the intensity and phase properties of the amplified OAM beams for the collinear case, and compare them with the input beams, as shown in Fig. (2). In the experiment, the pump light has a Gaussian-shape intensity distribution with the diameter of 2mm. To achieve full amplification, the diameter of the signal should be less than that of the pump, and hence is set to 1.5mm. Figure 2(a) and (b) depict the intensity profiles of input signal and amplified signal beams, and their interference patterns produced with the same reference plane beam, as shown in Fig. 2(c) and (d), respectively. We can note that the intensity and phase structure of amplified beam are mainly determined by input signal beam as mentioned above. In order to testify the property of low mode noise in SBA process, we restore the OAM signal to a Gaussian beam before and after the amplification by an inverse OAM transformation, respectively, and analyze their phase structures, as shown in Fig. 2(e), (f), (g) and (h). It can be seen that no phase singularities in the restored input or output signal beam, indicating that no obvious mode noise is introduced in SBA process as discussed above.

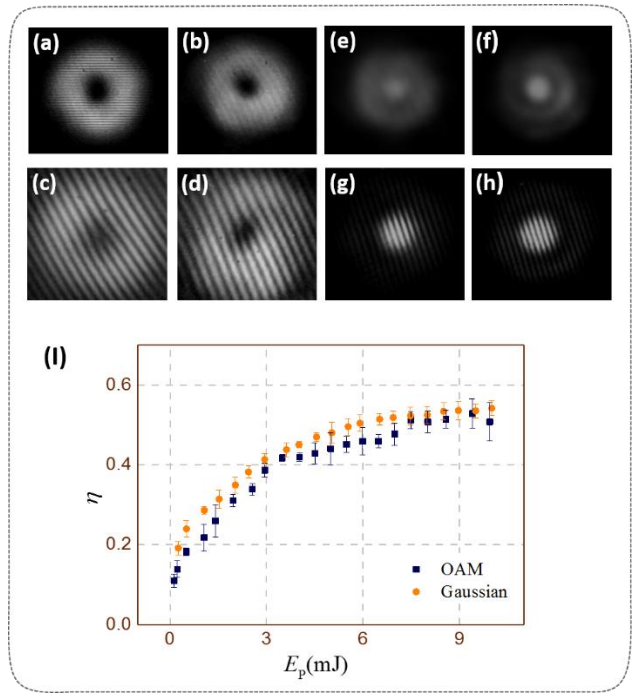
4

FIG. 2. (a)-(d) Near field patterns of input signal (a) and amplified signal (b), and their two-beam interferograms (c) and (d), respectively. (e)-(h) Near field patterns of restored input signal (e) and restored output signal (f), and their two-beam interferograms (g) and (h), respectively. (I) Energy conversion efficiency $\eta$ versus the pump energy $E_P$, including OAM signal and Gaussian signal.

We then measure the energy conversion efficiency ($\eta$) for different pump energies ($E_P$), the signal energy is chosen as 0.1mJ. In previous researches, we found that the photon-phonon coupling between pump and signal is much stronger than that between pump and random fluctuating phonon when the ratio of the signal intensity to the pump intensity ($I_{Sin}/I_P$) is greater than $10^{-4}$, and thus the self-SBS of the pump can be completely suppressed.[23] This point is also suitable for OAM signal in the same experimental conditions, for the presence of OAM does not modify the phase matching property. We can see that $\eta$ will exceed 40% for $E_P$ >4mJ regardless of signal mode, as described in Fig. 2(I), meanwhile, a clean OAM mode can be obtained as shown in Fig. 2(b). Therefore, the SBA process is an ideal mechanism for regulating and matching energy level between different physical or applied layers, especially for OAM-based integrated waveguide, and this mechanism can also be used in laser beam combination system to achieve high-power OAM mode output, satisfying the requirements of high energy applications.

Undeniably, a collinear frame can provide higher conversion efficiency and gain for phase matching and longer interaction length. However, from the Fig. 3(a) we can see that amplified signals will be covered by the self-SBS noise of pump for the case of very weak input signals[23], the self-SBS cannot be suppressed when signal energy is 0.1nJ. By contrast, in noncollinear frame, the amplified signal can be separated from self-SBS noise conveniently for their different propagating directions besides a simple configuration as shown in Fig. 3(b),[24-25] and then, we study the noncollinear case for OAM signals. Similar to the collinear experiment mentioned above, a vortex beam carrying an OAM of $\hbar$ per photon is used as the input Stokes signal, its intensity distribution and two-beam interferogram are still illustrated by Fig. 2(a) and 2(b), respectively. Variously, there is a small crossing angle $\theta$ between the pump and signal. Figure 3(c), (d), (e) and (f) show the intensity images and interferograms of amplified signals at the angle of 10mrad and 22mrad, respectively. It can be noted that the doughnut-shaped intensity profiles and single fork-shaped interference patterns of amplified signal are consistent with those of input vortex signals for two different angles. This further verifies that the mode of amplified beam is independent of the geometrical frames of photon-phonon coupling. Moreover, we achieve a high-gain amplification for a weak vortex signal, as seen in Fig. 3(g), a signal amplification factor (SAF) of $10^9$ is obtained when the vortex signal is reduced by about $10^{-12}$J, and this agrees with the results of OAM uninvolved.[25] The numerical difference between OAM signal and Gaussian signal in Fig. 2(I) and 3(g) is mainly because of the effect of SPP on signal beams.

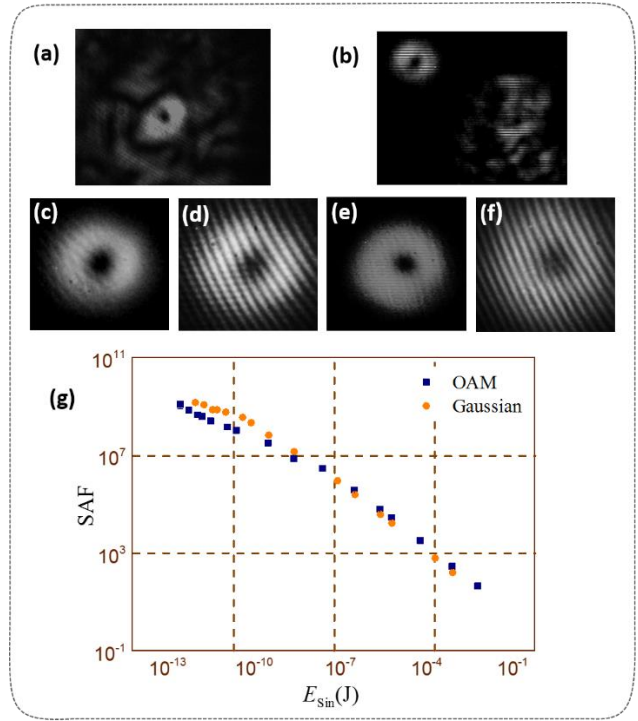

FIG. 3. (a), (b) Near field patterns of the amplified weak signal with self-SBS noise in collinear (a) and noncollinear (b) frames. (c)-(f) Near field patterns of amplified signal (c) and its two-beam interferogram at 10mrad (d), and those (e) and (f) at 22mrad, respectively. (g) Signal amplification factor (SAF) versus the signal energy $E_{Sin}$ at $\theta$=10mrad, including OAM signal and Gaussian signal.

In summary, we report a quasi-parametric amplification of OAM beams based on light-acoustic interaction. The results prove that SBA process is an ideal amplifying mechanism for OAM signals, in which the OAM beams can be efficiently amplified and the mode of output signal shows geometrical frame independent. Specifically, the process exhibits low noise level and no requiring extra pump sources compared with laser amplification and OPA, respectively. And this mechanism can be well applied in OAM fiber or integrated photonic-phononic waveguide, which shall be considered in the future. Beyond its advantage of mechanism in signal amplification, it should be further noted that the vortex periodical optical potential induced by vortex optical fields will exert a long-range interaction, i.e. optical torque, on the medium to excite vortex phonons. Mendonca et al. predicted that a well-defined phonon OAM can be excited from outside of media,[26] and our results suggest that the SBA involving OAM is a kind of feasible method. Furthermore, the OAM of excited phonons lies with the optical torque whose OAM changes with the OAM in optical fields and the geometrical frame of SBA, and this dynamics shows a large potential in OAM-multiplexing, especially for information manipulating layer.

**ACKNOWLEDGMENTS**

This work is supported by the National Natural Science Foundation of China (Grant No. 61378003), the Key Programs of the Natural Science Foundation of Heilongjiang Province of China (Grant No. ZD201415).